\documentclass{article}
\usepackage{url}
\usepackage{tikz}
\usepackage[]{algorithm}
\usepackage[noend]{algorithmic}
\usepackage{enumitem}
\usepackage{threeparttable}
\usepackage{float}
\usepackage[utf8]{inputenc}
\usepackage[english]{babel}
\usepackage{mathtools}
\usepackage{amsmath}
\usepackage{amsfonts}
\usepackage{comment}

\newtheorem{definition}{Definition}

\newcommand{\OO}{\mathcal{O}}
\newcommand{\ZKRP}{\mathrm{ZKRP}}
\newcommand{\ZKSM}{\mathrm{ZKSM}}

\newcommand{\SD}{\mathrm{SD}}
\newcommand{\LI}{\mathrm{LI}}
\newcommand{\NLI}{\mathrm{WT}}
\renewcommand{\SS}{\mathrm{SS}}
\renewcommand{\S}{\mathrm{S}}
\newcommand{\Setup}{\mathrm{Setup}}
\newcommand{\Prove}{\mathrm{Prove}}
\newcommand{\Verify}{\mathrm{Verify}}
\newcommand{\Commit}{\mathrm{Commit}}
\newcommand{\Open}{\mathrm{Open}}

\newcommand{\params}{\mathrm{params}}
\newcommand{\zkproof}{\mathrm{proof}}
\newcommand{\Prob}{\mathrm{Prob}}
\newcommand{\PK}{\mathrm{PK}}
\newcommand{\ZZ}{\mathbb{Z}}
\newcommand{\y}{\mathbf{y}}
\newcommand{\g}{\mathbf{g}}
\newcommand{\h}{\mathbf{h}}
\newcommand{\pp}{\mathbf{bp}}
\newcommand{\G}{\mathbb{G}}
\renewcommand{\a}{\mathbf{a}}
\renewcommand{\b}{\mathbf{b}}
\renewcommand{\l}{\mathbf{l}}
\newcommand{\rr}{\mathbf{r}}
\newcommand{\0}{\mathbf{0}}
\newcommand{\1}{\mathbf{1}}
\newcommand{\2}{\mathbf{2}}
\renewcommand{\k}{\mathbf{k}}
\newcommand{\Z}{\mathbb{Z}}
\newcommand{\rhs}{\mathrm{rhs}}
\newcommand{\IP}{\mathrm{IP}}
\newcommand{\RP}{\mathrm{RP}}
\newcommand{\MapToGroup}{\mathrm{MapToGroup}}

\newcommand{\Hash}{\mathrm{Hash}}
\newcommand{\Recursion}{\mathrm{ComputeProof}}
\newcommand{\CompGen}{\mathrm{ComputeGenerators}}
\newcommand{\out}{\mathrm{output}}
\newcommand{\commit}{\mathrm{commit}}
\newcommand{\s}{\mathbf{s}}
\newcommand{\p}{\mathbf{p}}
\newcommand{\meq}{\stackrel{?}{=}}
\newcommand{\myin}{\stackrel{?}{\in}}

\floatname{algorithm}{Algorithm}

\begin{document}

\tikzstyle{arrow} = [thick,->,>=stealth, line width=.5pt]

\title{A Survey on Zero Knowledge Range Proofs and Applications}
\author{Eduardo Morais \and Tommy Koens \and Cees van Wijk \and Aleksei Koren}
\date{3 December 2018}

\maketitle

\begin{abstract}
\noindent In last years, there has been an increasing effort to leverage Distributed Ledger Technology (DLT), including blockchain. One of the main topics of interest, given its importance, is the research and development of privacy mechanisms, as for example is the case of Zero Knowledge Proofs (ZKP). ZKP is a cryptographic technique that can be used to hide information that is put into the ledger, while still allowing to perform validation of this data. 

In this work we describe different strategies to construct Zero Knowledge Range Proofs (ZKRP), as for example the scheme proposed by Boudot~\cite{boudot} in 2001; the one proposed in 2008 by Camenisch et al~\cite{ccs08}, and Bulletproofs~\cite{bulletproofs}, proposed in 2017. We also compare these strategies and discuss possible use cases. 

Since Bulletproofs~\cite{bulletproofs} is the most efficient construction, we will give a detailed description of its algorithms and optimizations. Bulletproofs is not only more efficient than previous schemes, but also avoids the trusted setup, which is a requirement that is not desirable in the context of Distributed Ledger Technology (DLT) and blockchain. In case of cryptocurrencies, if the setup phase is compromised, it would be possible to generate money out of thin air. Interestingly, Bulletproofs can also be used to construct \emph{generic} Zero Knowledge Proofs (ZKP), in the sense that it can be used to prove generic statements, and thus it is not only restricted to ZKRP, but it can be used for any kind of Proof of Knowledge (PoK). Hence Bulletproofs leads to a more powerful tool to provide privacy for DLT\@. Here we describe in detail the algorithms involved in Bulletproofs protocol for ZKRP. Also, we present our implementation, which was open sourced~\cite{impl}. 
\end{abstract}

\newpage

\section{Introduction}
\label{int}

DLT and blockchain have been subject to intense research in last years, because it allows to construct consensus among parties that do not fully trust each other, without the necessity of a trusted third party. However, in public and permissionless ledgers, transactions can be viewed by everyone in the network. This fact is a hindrance that we must overcome if those transactions contain privacy-sensitive information.

In order to protect private information, a possible alternative is to use a Trusted Execution Environment (TEE), like Intel SGX~\cite{tommysgx} technology. The idea is that any private data must appear in the blockchain in encrypted form. Only the owners of the subjacent cryptographic keys will be able to decrypt it. Validation of this information must be done in the TEE system, where the cryptographic keys can be embedded. Therefore, private data will only be visible after decryption, which occurs inside a controlled environment. Putting differently, a TEE offers protection against information leakage by restricting manipulation of private data to a region of memory that can not be accessed by other processes in the same machine, or even by its administrator. Nevertheless, attacks~\cite{sgxattack2,sgxattack1} to SGX where proposed in literature, showing that this technology is vulnerable to \emph{branch prediction} and \emph{side-channel} attacks, respectively. 

A different approach to secure private data is ZKP, which is a cryptographic technique that have been used to provide \emph{privacy by design} in the context of DLT and blockchain. Shortly, ZKP allows an entity called \emph{prover} to argue to another party, called \emph{verifier}, that a determined statement is true without revealing more information than strictly necessary to convince her.  

In previous works~\cite{ing1,ing2} ING described some preliminary results. The purpose of this work is to extend them in order to provide a complete survey on ZKRP protocols. 

In summary, ZKRP allows to prove that a secret integer belongs to a certain interval. For example, if we define this interval to be all integers between 18 and 200, a person can use the ZKRP scheme to prove that she is over 18. This gives her permission, according to some regulation, to consume a determined service, but without revealing her specific age. In the context of payment systems, if party $A$ wants to transfer money to party $B$, then it is possible to utilize ZKRP to prove that the amount of money in the transaction is positive, otherwise, if the amount is negative, such transaction would in fact transfer money in the opposite direction, i.e. from $B$ to $A$. A limitation of ZKRP is, however, that it can be used for numeric intervals only, and it is not possible to use a generic set. 

In this document we will describe some strategies to achieve ZKRP. In particular, we will describe the construction by Camenisch et al~\cite{ccs08}, which allows to construct \emph{Zero Knowledge Set Membership} (ZKSM). With ZKSM we can define generic sets and still maintain privacy requirements. ZKSM is very similar to ZKRP, the difference is that instead of the numeric interval used in ZKRP, we have a \emph{generic set} in ZKSM. In other words, imagine that the set is formed by all countries in the European Union. Hence, if the private information is given by a country name, for instance the country of residence of a particular user, then she can use ZKSM to generate a zero knowledge proof that the private data is indeed an element from this set, therefore proving that she lives in the EU. This kind of cryptographic building block is interesting for any situation that includes sets and includes a strong privacy component. More concretely, next we describe possible use cases for ZKRP and ZKSM. 

In the following sections we describe in detail the algorithms necessary to implement ZKRP and instantiate the underlying parameters in order to obtain an appropriate level of security. We also compare the different schemes with regards to proof size, and the complexity of the prover and verifier algorithms. 

\subsection{Contributions}

There are many surveys about zero knowledge proofs, but mostly related to the theoretical foundations of the proposed cryptographic constructions. The main goal of this survey is to bridge the gap between those papers whose audience is the cryptographic community and the community of developers that are more focused on implementation aspects. In 2018 there was the first Zero Knowledge Proofs Standardization Workshop~\cite{zkproof}, where academy and industry started the effort to produce a standard to implement ZKPs. The workshop was divided into three categories: security, implementation and application. The first one was responsible for establishing the subjacent theoretical basis to instantiate ZKPs, determining security models and underlying assumptions. The second one was responsible to propose APIs, software architecture and best practices for ZKPs. Lastly, the third one was responsible for determining interesting use cases for ZKPs, finding high level requirements to the other categories. In particular, some \emph{ZK gadgets} were identified as important building blocks for the construction of solutions to more complex problems. Among the ZK gadgets that were discussed, we can remark ZK Range Proofs, ZK Set Membership and cryptographic accumulators. These gadgets can be commonly used to solve different practical problems, as pointed out in last section.   

Next we summarize the main contributions of this work:

\begin{itemize}
\item Survey possible use cases for ZKRP and other similar ZKPs. We indicate which papers in the literature present important contributions for the construction of efficient solutions to this use cases.  
\item Describe in detail the algorithms required for each different strategy to implement ZKRP. In particular, we describe how the Fiat-Shamir must be implemented in order to obtain non-interactive protocols. 
\item Include Bulletproofs in the comparison presented in the work by Canard et al~\cite{survey1} and present our open source implementation~\cite{impl}.
\end{itemize}

\subsection{Organization}

In Section~\ref{appsec} we describe possible use cases for ZKRPs. In Section~\ref{fundsec} we give fundamental results that are important to understand the rest of the document. In Section~\ref{main} we describe in detail how to implement ZKRP using different strategies. In Section~\ref{impl} we describe our implementation, while in Section~\ref{comp} we compare the schemes with respect to proof size, prover and verifier complexities. In Section~\ref{fmsec} we discuss related work and give some final remarks. 

\section{Applications}
\label{appsec}

In order to give the reader a motivation to investigate further on Zero Knowledge Proofs, we present in this section some interesting applications. 

\begin{itemize}[label={--}]
	\item \textbf{Over 18.} ZKRP is a special case of ZKSM, due to the fact that any numeric interval is also a set. Therefore if the ZKSM is more efficient than ZKRP, what turns out to be true in certain scenarios, then ZKSM can replace ZKRP to improve performance.  
	\item \textbf{KYC.} As explained above, ZKSM allows to validate that a determined piece of private information belongs to a set of valid values. This property may be used to ensure compliance, while preserving a client's privacy. For example, an interesting use case is the so-called \emph{anonymous credentials}, where a trusted party can attest that a user credential contains attributes whose values are correct, namely the country of residence of a person being validated by government, allowing the user to later prove that she lives in a country that belongs to the European Union, without revealing which country.
	\item \textbf{Electronic voting.} This is an important topic of research, which attracted the attention of many researchers in last years. Different solutions~\cite{DJN2010,gro05,BGN,Helios} were proposed to different types of elections. Some solutions are based on zero knowledge proofs, like ZKRP, proof of shuffling, proof of decryption and other related techniques, while others use different cryptographic primitives, like homomorphic threshold encryption and Multi-Party Computation (MPC). 
	\item \textbf{Electronic auctions and procurement.} Secure electronic auctions is a subject that has being focus of research for a long time~\cite{lan}, and it is an important motivation to in the study of ZKRPs, since it is one of the main cryptographic techniques that can be used to construct secure protocols.  In particular, it is possible to remark the proposal of secure constructions~\cite{Rab06a,Rab06b,MR14} for Vickrey auctions, where the winner pays the second highest bid. A complementary problem to electronic auctions is \emph{procurement}, where parties concur for the lowest price. According to the World Bank report~\cite{WBrep}, the volume of bribes in public sector procurement is roughly US\$200 billion per year.  
	\item \textbf{Board membership.} ZKSM can be used to construct \emph{ring signatures}, which allows someone to digitally sign a message in behalf of a group of users. Afterwards, anyone can verify the signature indeed was generated by a member of the group. This is interesting for example to allow a member of a directing board to anonymously sign a contract. 
	\item \textbf{Anti-Money Laundering (AML).} If we define the ZKSM set to be a list of entities that can consume a determined service, then we can construct a \emph{whitelist} and an anonymous entity can prove that it is whitelisted and thus has permission to use that service. Similarly, it is possible to construct a \emph{blacklist} formed by criminals, or by countries that are considered to be non-cooperative against money laundering, as is the case of the Financial Action Task Force~\cite{faft} (FATF) blacklist. Hence an anonymous entity can prove that it does not belong to the blacklist, ensuring AML compliance.  
	\item \textbf{Reputation validation.} Consider a set formed by companies that have good reputation, either because they are compliant to some regulation or due to the fact that they are good payers, or, in general, because they respect certain conditions. Then it is possible to use ZKSM to produce a \emph{proof of reputation}. This use case is a little bit different compared to the previous ones, since in many practical scenarios we can not make public the set of companies that have reputation or not. In other words, this set itself is private. In this case, we must have a solution that is a little bit different from the construction presented here in the paper. Actually, there is line of research devoted to this topic, which is called \emph{cryptographic accumulators}. Although accumulators can not be directly constructed based on ideas presented here in this document, there is indeed a close relation between ZKSM and accumulators. In fact, one of the authors of the ZKSM paper~\cite{ccs08} described here, namely Camenisch, has many papers~\cite{accumulators1,accumulators2,accumulators3} in this area.  
	\item \textbf{Common Reporting Standard (CRS).} In 2014 forty-seven countries agreed on the CRS proposal~\cite{crs}, whose main goal is to provide \emph{transparency} in a global level regarding financial information, in particular to avoid tax fraud and tax evasion. The CRS allows automatic exchange of information, based on XML schemas that are responsible to dynamically describe the data format and validation patterns. Using ZKSM it is possible to carry on some of those possible validations, such as enumerations and integer ranges. Hence we have that private-sensitive data can be validated even if it is sent in its encrypted form. Therefore ZKSM may be considered an important tool that can be \emph{reused} to provide \emph{privacy on demand}.  
\end{itemize}

The applications described above are general purpose, but could be interesting also in the context of DLT and blockchain technology. Next we focus on application that are important in the specific scenario of DLT and blockchain:

\begin{itemize}[label={--}]
\item \textbf{Confidential Transactions and Mimblewimble.} In 2016, Confidential Transactions (CT) were proposed by Maxwell~\cite{ct}, which utilizes Pedersen commitments~\cite{pedersen} to \textbf{hide transactions amounts}. Instead of publishing the amounts being spent in the clear, each party uses the commitment scheme to hide the amount, what makes it infeasible for an adversary to obtain any information about transaction denominations. Since a Pedersen commitment is homomorphic, it allows transactions outputs to be added up without requiring to open the subjacent commitments. Also, the commitment can be used to generate a ZKRP, which is sufficient to validate that a transaction is correct. For instance, it is necessary to show that the amount lies in the interval $[0, 2^n)$, where $2^n$ is considerably smaller than the size of the underlying group used to construct the Pedersen commitment, ensuring there is no overflow; and $2^n$ is big enough to deal with every possible valid denomination. 

However, the usage of ZKRP would make the size of transactions too big. Namely, CT with just two outputs and 32 bits of precision would require roughly a ZKRP whose size is 5 KB, leading to transactions whose total size is equal to 5.4 KB\@. Thus, ZKRP would correspond to almost 93\% of the transaction size. Therefore in order to use CT in Bitcoin, we would need 160 GB only for ZKRP\@. If Bulletproofs where used in replacement of the underlying range proof used in CT, then it would reduce this requirement to only 17 GB. 

Mimblewimble~\cite{mimblewimble} is an optimization to CT that can make the size of the ledger even smaller, by aggregating and compressing transactions in such a way that avoids the necessity to download old and unspent transactions outputs.  

\item \textbf{Provisions.} Provisions~\cite{provisions} is a protocol that allows a Bitcoin exchange to prove it is \textbf{solvent}, by showing that each account has positive balance, and also showing that the exchange has an amount of funds that is larger than or equal to the summation of all individual account's balance in the system. The challenge here is to calculate a single zero knowledge proof based on the information provided by different participants. This is difficult because each individual balance is encrypted using distinct keys, thus combining them is not straightforward, and requires MPC. Bulletproofs has a MPC protocol that solves this problem efficiently. For instance, if we consider a cryptocurrency exchange with 2 million clients, current implementation of Provisions requires 62 MB of ZKRPs. However, using Bulletproofs this number can be reduced to less than 2 KB, which corresponds to an optimization factor of 300. 

\item \textbf{Private smart contracts.} Ethereum~\cite{ethereum} allows to construct \emph{smart contracts} over blockchain, which can be seen as generic applications running in a distributed way, therefore avoiding the necessity to have a centralized solution. In other words, a smart contract is a piece of code that will run by all participants in Ethereum network. However, since there is no mechanism to provide privacy to the system, we have that all the information in the smart contracts is visible by every other party, what constitutes a huge issue in many scenarios\@. This problem could be solved by using zk-SNARKs~\cite{zksnarks}, but it requires a trusted setup, and this problem is even worse in the case of smart contracts, because we need a new setup for each contract. Hawk~\cite{hawk} is an interesting proposal to implement private smart contracts, however it not only needs a new setup for each contract, but also requires a \emph{trusted manager}, who can view the user private information. Bulletproofs is an interesting proposal regarding private smart contracts, since it avoids the trusted setup and offers a generic ZKP protocol which has small proofs. 
\end{itemize}

\section{Fundamentals}
\label{fundsec}

In this section we define commitment schemes, zero knowledge proofs and other important components that are necessary in order to comprehend this work. The purpose of this section is not to present very formal definitions. To achieve this goal, the reader can use Goldreich's book~\cite{goldreich}.  

\textbf{Notation.} Notation $x \in_R S$ is used when variable $x$ is set to a random element of set $S$. We are going to use Camenisch and Stadler~\cite{stadler} notation for proofs of knowledge: 

$$\PK\{(\delta, \gamma) : y = g^\delta h^\gamma \wedge (u \leq \delta \leq v)\},$$
which denotes a proof of knowledge of integers $\delta$ and $\gamma$ such that $y = g^\delta h^\gamma$ and $u \leq \delta \leq v$. In other words, this notation means that $y$ is the commitment to the secret value $\delta$, which is contained in the interval $[u, v)$. Greek letters are used to denote values that must be known only to the prover. For instance, we have that $\delta$ is her private data, while $\gamma$ is a random value that is used to hide $\delta$.

Finally, we use notation $x \meq y$ to check if $x$ is equal or not to $y$. 

\subsection{Assumptions}

The constructions presented in this paper are based on the assumptions described in this section. 

The strong RSA assumption first appeared in the work of Fujisaki and Okamoto~\cite{FO}. It is a stronger assumption with respect to the conventional RSA assumption, because any adversary who can break the RSA assumption would also be able to break the strong RSA assumption. 

\begin{definition}
    \textbf{Strong RSA assumption.} Given an RSA-modulus $n$ and an element $y \in \ZZ_n^\star$, it is infeasible to find integers $e \neq \pm 1$ and $x$, such that $y = x^e \pmod{n}$. 
\end{definition}

\begin{definition}
    \textbf{Discrete Logarithm assumption.} Let $\G$ be a group of prime order $q$, a generator $g \in \G$ and an arbitrary element $y \in \G$, it is infeasible to find $x \in \ZZ_q$, such that $y = g^x$. 
\end{definition}

\begin{definition}
\label{qstrongdh}
    \textbf{$q$-Strong Diffie-Hellman assumption.} Given groups $\G_1$ and $\G_T$, associated with a secure bilinear pairing map $e$; given generator $g \in \G_1$ and powers $g^x, \dots, g^{x^q}$, for $x \in_r \ZZ_p$, we have that it is infeasible for an adversary to output $(c, g^{1/(x+c)})$, where $c \in \ZZ_p$. 
\end{definition}

It is important to remark that these assumptions are not valid if quantum computers come to existence. Therefore, the research of quantum-resistant ZKPs is a very important subject. 

\subsection{Commitment}

Shortly, a cryptographic commitment allows someone to compute a value that hides some message without ambiguity, in the sense that no one later will be able to argue that this value corresponds to a different message. In other words, given the impossibility to change the hidden message, we say that the user committed to that message. The purpose of using a commitment scheme is to allow a prover to compute zero knowledge proofs where the hidden message is the underlying witness $w$.

\begin{definition}
A \textbf{commitment scheme} is defined by algorithms $\Commit$ and $\Open$ as follows:
	
\begin{itemize}[label={--}]
	\item $c = \Commit(m, r)$. Given a message $m$ and randomness $r$, compute as output a value $c$ that, informally, hides message $m$ and such that it is hard to compute message $m'$ and randomness $r'$ that satisfies $\Commit(m', r') = \Commit(m, r)$. In particular, it is hard to invert function $\Commit$ to find $m$ or $r$. 

	\item $b = \Open(c, m, r)$. Given a commitment $c$, a message $m$ and randomness $r$, the algorithm returns true if and only if $c = \Commit(m, r)$.
\end{itemize}
\end{definition}

A commitment scheme has 2 properties:

\begin{itemize}
	\item \textbf{Binding.} Given a commitment $c$, it is hard to compute a different pair of message and randomness whose commitment is $c$. This property guarantees that there is no ambiguity in the commitment scheme, and thus after $c$ is published it is hard to open it to a different value.  
	\item \textbf{Hiding.} It is hard to compute any information about $m$ given $c$.
\end{itemize}

A well known commitment scheme is called \emph{Pedersen commitment}~\cite{pedersen}. Given group $\ZZ_p$, of prime order $p$, where the discrete logarithm problem is infeasible, the commitment is computed as follows:  

$$c = \Commit(m, r) = g^m h^r.$$

In order to open this commitment, given message $m$ and randomness $r$, we simply recompute it and compare with $c$. An interesting property is that Pedersen commitment is \emph{homomorphic}. Namely, we have that for arbitrary messages $m_1$ and $m_2$ and randomness $r_1$ and $r_2$, such that $c_i = \Commit(m_i, r_i)$ for $i \in \{1,2\}$, then

$$c_1.c_2 = \Commit(m_1 + m_2, r_1 + r_2).$$ 

Pedersen commitment is commonly implemented using groups over elliptic curves instead $\ZZ_p$. Also, it is important to remark that if the discrete logarithm of $h$ with respect to $g$ is known, then it is easy to generate $m'$ and $r'$ such that $\Commit(m', r') = \Commit(m, r)$, breaking the binding property. Thus in order to generate $h$ securely, we must use a hash function that maps binary public strings to elliptic curve points~\cite{maptogroup}. 

Another commitment scheme that will be required later in this document is the Fujisaki-Okamoto commitment~\cite{FO}. The formula to calculate the commitment itself is the same as in Pedersen commitment, namely $g^m h^r$. The difference is the underlying group, which for the Fujisaki-Okamoto is given by an RSA group $\ZZ_n$, where $n=pq$ and $p$ and $q$ are \emph{safe primes}, what means that $(p-1)/2$ and $(q-1)/2$ are also prime numbers. Also, we have that the domain over which randomness $r$ is chosen is different, because the Fujisaki-Okamoto commitment requires $r \in [2^{-s}n+1,2^s n - 1]$, with $s$ chosen in such a manner that $2^{-s}$ is negligible. Interestingly, in the original paper~\cite{FO} Fujisaki and Okamoto propose an interactive protocol for Zero Knowledge Range Proofs, but unfortunately the performance is not good for practical usage.   

\subsection{Zero Knowledge Proofs}

Zero Knowledge Proofs (ZKP) were proposed in 1989 by Goldwasser, Micali and Rackoff~\cite{gmr89}. Using this kind of cryptographic primitive it is possible to show that some statement is true about a secret data, without revealing any other information about the secret beyond this statement. Since then, ZKP became an important field of research, because it provides a new characterization of the complexity class NP, using the so-called \emph{interactive programs}, and also because it is very useful to construct many cryptographic primitives. Given an element $x$ of a language $\mathcal{L} \in NP$, an entity called \emph{prover} is able to convince a verifier that $x$ indeed belongs to $\mathcal{L}$, i.e.\ there exists a witness $w$ for $x$. In particular we are interested in \emph{proof of knowledge} (PoK), where the prover not only convinces about the existence of some witness, but also shows that the prover in fact knows a specific witness $w$. A desirable characteristic of such proof systems is \emph{succinctness}, informally meaning that the proof size is small and thus can be verified efficiently. Such constructions are called zk-SNARKs~\cite{zksnarks}. However, although asymptotically good, zk-SNARKs still have some limitations and for some specific problems it turns out that different approaches achieve better performance, as we will show in this document.

Nowadays ZKP is being used to provide privacy to DLT and blockchain. For instance, it allows to design private payment systems. In summary, we would like to permit parties to transfer digital money, while hiding not only their identities but also the amount being transferred, known as \emph{denomination}. ZKP can be used to hide this information, but still permitting validation of transactions. An important validation is showing that the denomination is positive, otherwise some payer would be able to receive money by using negative amounts. In this context we have that zk-SNARKs don't provide good performance when compared to protocols designed specifically for this purpose. The focus of this document is the description of different constructions of ZKRP and compare them to understand when to use each scheme in practice. More concretely, ZKRP allows some party Alice, known as the \emph{prover}, and who possesses a secret $\delta$, to prove to another party Bob, known as the \emph{verifier}, that $\delta$ belongs to the interval $[u, v)$, for arbitrary integers $u$ and $v$. 

\begin{definition}
	A \textbf{Non-Interactive Zero Knowledge (NIZK) proof} scheme is defined by algorithms $\Setup$, $\Prove$ and $\Verify$ as follows:  

	\begin{itemize}[label={--}]
	\item $\Setup$ algorithm is responsible for the generation of parameters. Concretely, we have that $\params  = \Setup(\lambda)$, where the input is the security parameter $\lambda$ and the output is the parameters of the ZKP system of algorithms. 

	\item $\Prove$ syntax is given by $\zkproof = \Prove(x, w)$. The algorithm receives as input an instance $x$ of some NP-language $\mathcal{L}$, and the witness $w$, and outputs the zero knowledge proof. 

	\item $\Verify$ algorithm receives the proof as input and outputs a bit $b$, which is equal to $1$ if the verifier accepts the proof. 
\end{itemize}
\end{definition}

It is important to remark that not all ZKP schemes are non-interactive. On contrary, most ZKP protocols described in the literature are in fact interactive. In general, the prover must answer \emph{challenge messages} sent by the verifier in order to convince him that the proof is valid, what requires multiple rounds of communication. In the context of DLT and blockchain  applications, we would like to avoid this communication, because either (i) validating nodes can not properly agree on how to choose those challenges, since in many constructions we have to choose them randomly, while the verification algorithm must be deterministic in order to reach consensus; or (ii) because it would make the communication complexity of the system very poor. Nevertheless, the Fiat-Shamir heuristic~\cite{fs} is a generic technique that allows to convert interactive ZKP schemes into non-interactive protocols. The drawback of this heuristic is that it makes the cryptosystem secure under the \emph{random oracle model}~\cite{rom} (ROM). In particular, it is straightforward to make the ZKRP schemes described in this document non-interactive using the Fiat-Shamir heuristic.  

A zero knowledge proof scheme has the following properties:

\begin{itemize}
	\item \textbf{Completeness.} Given a witness $w$ that satisfies instance $x$, we have that $\Verify(\Prove(x,w)) = 1$.

	\item \textbf{Soundness.} If the witness $w$ does not satisfy $x$, then the probability $\Prob[\Verify(\Prove(x,w)) = 1]$ is sufficiently low. 

	\item \textbf{Zero Knowledge.} Given the interaction between prover and verifier, we call this interaction a \emph{view}. In order to capture the zero knowledge property we use a polynomial-time \emph{simulator}, which has access to the same input given to the verifier (including its randomness), but no access to the input of the prover, to generate a \emph{simulated view}. We say that the ZKP scheme has \emph{perfect zero knowledge} if the simulated view, under the assumption that $x \in \mathcal{L}$, has the same distribution as the original view. We say that the ZKP scheme has \emph{statistical zero knowledge} if those distributions are \emph{statistically close}. We say that the ZKP scheme has \emph{computational zero knowledge} if there is no polynomial-time distinguisher for those distributions. Intuitively, the existence of such a simulator means that whatever the verifier can compute from the interaction with the prover, it was already possible to compute before such interaction, hence the verifier learned nothing from it. Also, we say that it is a \emph{proof of knowledge} if we can find an \emph{extractor}, who has \emph{rewindable} black-box access to the prover, that can compute the witness $w$ with non-negligible probability.  
\end{itemize}

\subsection{Bilinear Pairings}

Some constructions of ZKRP are based on the existence of a secure bilinear map $\pp = (\G_1, \G_2, \G_t, e, g_1, g_2)$, where $\G_1$, $\G_2$ and $\G_t$ are groups of sufficiently large prime order, $g_1$ and $g_2$ are generators of $\G_1$ and $\G_2$ respectively and $e$ is an appropriate choice of bilinear map, satisfying the usual requirements: (i) non-degeneracy; (ii) efficiently computable and (iii) bilinearity. This cryptographic primitive is key to the constructions we will present in the next sections and it is important to remark that care must be taken when instantiating such primitive~\cite{DummyPairings,pairings}. Barreto-Naehrig~\cite{bn} elliptic curves permit to implement bilinear maps efficiently.

\section{Zero Knowledge Range Proofs}
\label{main}

The first constructions of ZKRP protocols were presented decades ago, with schemes like the one proposed in 1995 by Damg{\aa}rd~\cite{Damgard95} and in 1997 by Fujisaki and Okamoto~\cite{FO}. Unfortunately those proposals are not efficient to be used in practice. The first practical construction was proposed by Boudot in 2001~\cite{boudot}. In this document we will focus on constructions that came after Boudot's proposal. 

In this section we describe in detail different strategies to achieve ZKRP. A summary of the main characteristics of each family of constructions follows:  

\begin{itemize} 
	\item \textbf{Square decomposition.} One of the ideas that can be used to obtain zero knowledge range proofs is the decomposition of the secret element into a sum of squares, as proposed in 2001 by Boudot~\cite{boudot}. In 2003 Lipmaa et al~\cite{lip03} improved the construction using Lagrange's \emph{four squares theorem}. In 2005 Groth~\cite{gro05} observed that if the element is in the form $4n+1$, then it is possible to get the same result by decomposing only into three squares. The drawback of this approach is that the algorithm by Rabin and Shallit~\cite{RS}, required for the decomposition into squares, runs in time $\OO(k^4)$, where $k$ is the size of the secret. Both Lipmaa~\cite{lip03} and Groth~\cite{gro05} improved this algorithm, but in practice we have that it leads to a poor performance for the Prover's algorithm. 

\item \textbf{Signature-based.} Another idea for the prover is to prove, in a blind way, that he knows a signature on the secret. Initially, all elements in the interval are signed, then the proof that the prover knows the signature means that this integer belongs to the expected interval. In fact this interval can be any possible finite set, which means that this solution can be used to construct ZK Set Membership. In 2008 Camenisch, Chaabouni and shelat used bilinear pairings to construct an efficient ZKSM scheme~\cite{ccs08} that may be used also for ZKRP.

\item \textbf{Multi-base decomposition.} A common approach that one could follow to build ZKRP schemes is to decompose the secret into the bit representation, which allows to prove that it belongs to the interval by using Boolean arithmetic. Basically, the prover must commit to each bit of the secret; provide a zero knowledge proof that it is indeed a bit; and show a zero knowledge proof that the representation is valid. This last condition may easily be achieved by the utilization of homomorphic commitments. If instead of using the bit representation we use $u$-ary representation, then we can obtain more efficient constructions, as pointed out in~\cite{ccs08}. Another possible strategy is to use the so-called \emph{multi-base decomposition}~\cite{bin01,mbase03}, which is an alternative way to represent the secret and it allows to build ZKRP schemes that are good for the case of small secrets. In~\cite{survey1} the authors propose a new scheme and provide a comparison among different proposals in the literature. In summary, regarding the verifier's complexity, their construction is good for very small secrets (5 bits). For secret bit-length between 5 and 25 the scheme proposed in~\cite{mbase03} is the best option, while for more than 25 bits the signature-based scheme proposed in~\cite{ccs08} is the best alternative. It is important to remark that although the square decomposition strategy has constant complexity, it only provides an interesting performance for huge secrets, say, more than 500 bits. The reason is because those schemes are based on the strong RSA assumption, which requires big variables and inherently allows the usage of big secrets.    

\item \textbf{Bulletproofs.} Unfortunately, all the schemes previously above-mentioned depends upon a \textbf{trusted setup}, which may not be interesting in the context of cryptocurrencies. For instance, if an adversary is able to circumvent this trusted setup, he would be able to create money out of thin air. Recently, B\"unz et al~\cite{bulletproofs} proposed a new idea to construct ZKRP, which they called \emph{Bulletproofs}. They proposed to use an inner product proof in order to achieve ZKRP with very small proof sizes. Also, they showed how to use a component called \emph{multi-exponentiation} in order to optimize their construction. The authors also provided an efficient implementation that shows their proposal is adequate for many practical scenarios. However, this proposal was not included in the comparison by Canard et al~\cite{survey1}, then one of the contributions of this work is to analyze how Bulletproofs compares to the other proposals. Actually, Bulletproofs is based on the decomposition of the secret into its bit representation, thus it fits the category described previously, but since it proposed a different framework for generating ZKPs, namely using PoK for inner product relations, then we considered important to separate Bulletproofs in order to give it more focus. 
\end{itemize}

\subsection{Square decomposition construction}
\label{boudot}

In this section we describe the algorithms necessary to implement the ZKRP proposed by Boudot~\cite{boudot} in 2001. This construction requires some building blocks, like the zero knowledge proof that two commitments hide the same secret and the zero knowledge proof that the secret is a square. These schemes are based on the \emph{strong RSA assumption}, then we have that $n$ must be the product of \emph{safe primes}. 

We denote the zero knowledge proof that two commitments hide the same secret by $\PK_\SS = \{x, r_1, r_2 : E = g_1^x h_1^{r_1} \wedge F = g_2^x h_2^{r_2}\}$. The parameters for the $\PK_\SS$ scheme is given by $\params_\SS = (t, \ell, s_1, s_2)$, which must be set in order to achieve the desired level of security. Namely, we have that soundness is given by $2^{t-1}$, while the zero knowledge property is guaranteed given that $1/\ell$ is negligible. Next we present algorithms $\Prove_\SS$ and $\Verify_\SS$. It is important to remark that the discrete logarithm of $g_1$ with respect to $h_1$, or its inverse, must be unknown, otherwise the commitment is not secure. Analogously, we have that the same condition must be valid for $g_2$ and $h_2$. The hash function is such that it outputs $2t$-bit strings. Finally, we have that $s_1$ and $s_2$ must be chosen in order to have secure commitments, i.e. $2^{s_i}$ must be negligible for $i \in \{1,2\}$.

\begin{algorithm}
\caption{Proof of Same Secret: $\Prove_{\SS}$}
\label{pssalg}
\begin{algorithmic}
\REQUIRE $x, r_1, r_2, E, F, \params_\SS$.
\ENSURE $\zkproof_\SS$.
\STATE $\omega \in_R [1,2^{\ell + t}b-1],$
\STATE $\eta_1 \in_R [1,2^{\ell + t + s_1}n-1],$
\STATE $\eta_2 \in_R [1,2^{\ell + t + s_2}n-1],$
\STATE $\Omega_1 = g_1^\omega h_1^{\eta_1},$
\STATE $\Omega_2 = g_2^\omega h_2^{\eta_2},$
\STATE $c = \Hash(\Omega_1 || \Omega_2),$
\STATE $D = \omega + cx,$
\STATE $D_1 = \eta_1 + cr_1,$
\STATE $D_2 = \eta_2 + cr_2,$
\RETURN $\zkproof_\SS = (c, D, D_1, D_2)$.
\end{algorithmic}
\end{algorithm}

\begin{algorithm}
\caption{Proof of Same Secret: $\Verify_{\SS}$}
\label{vssalg}
\begin{algorithmic}
\REQUIRE $E, F, \zkproof_\SS$.
\ENSURE True or false. 
\RETURN $c \meq \Hash(g_1^D h_1^{D_1}E^{-c} || g_2^D h_2^{D_2}F^{-c})$.
\end{algorithmic}
\end{algorithm}

We denote the zero knowledge proof that a secret is a square by $\PK_\S = \{x, r_1 : E = g^{x^2}h^r\}$. We have that $\params_\S = (t, \ell, s)$ represents the parameters for the $\PK_\S$ scheme, so that soundness is given by $2^{t-1}$ and the zero knowledge property is guaranteed if $1/\ell$ is negligible, as before. Algorithms~\ref{psalg} and~\ref{vsalg} corresponds to $\Prove_\S$ and $\Verify_\S$, respectively. Also, the discrete logarithm of $g$ with respect to $h$, or its inverse, must be unknown, otherwise the commitment is not secure. 

\begin{algorithm}
\caption{Proof of Square: $\Prove_\S$}
\label{psalg}
\begin{algorithmic}
\REQUIRE $x, r_1, E, \params_\S$.
\ENSURE $\zkproof_\S$.
\STATE $r_2 \in_R [-2^s n + 1, 2^s n - 1]$,
\STATE $F = g^x h^{r_2}$,
\STATE $r_3 = r_1 - r_2x$, 
\STATE $\zkproof_SS = \Prove_SS(x, r_2, r_3, E, F)$,
\RETURN $\zkproof_S = (E, F, \zkproof_\SS)$.
\end{algorithmic}
\end{algorithm}

\begin{algorithm}
\caption{Proof of Square: $\Verify_\S$}
\label{vsalg}
\begin{algorithmic}
\REQUIRE $\zkproof_\S$.
\ENSURE True or false.
\RETURN $\Verify_\SS(E, F, \zkproof_\SS)$.
\end{algorithmic}
\end{algorithm}

We denote the zero knowledge proof that a secret belongs to a larger interval, originally proposed by Chan et al~\cite{cft}, by using notation $\PK_\LI = \{x, r : E = g^x h^r \wedge x \in [-2^{t+\ell}b, 2^{t+\ell}b] \}$. We have that $\params_\LI = (t, \ell, s)$ represents the parameters for the $\PK_\LI$ scheme, so that completeness is achieved with probability greater than $1-2^\ell$; soundness is given by $2^{t-1}$ and the zero knowledge property is guaranteed if $1/\ell$ is negligible. Algorithms~\ref{plialg} and~\ref{vlialg} corresponds to $\Prove_\LI$ and $\Verify_\LI$, respectively. Also, the discrete logarithm of $g$ with respect to $h$, or its inverse, must be unknown. 

\begin{algorithm}
\caption{Proof of Larger Interval: $\Prove_\LI$}
\label{plialg}
\begin{algorithmic}
\REQUIRE $x, r, E, \params_\LI$.
\ENSURE $\zkproof_\LI$.
\REPEAT
\STATE $\omega \in_R [0, 2^{t+\ell}b-1]$,
\STATE $\eta \in_r [-2^{t+\ell+s}n+1, 2^{t+\ell+s}n-1]$,
\STATE $\Omega = g^\omega h^\eta \pmod{n}$,
\STATE $C = \Hash(\Omega)$,
\STATE $c = C \pmod{2^t}$,
\STATE $D_1 = \omega + xc$,
\STATE $D_2 = \eta + xc \in \ZZ$,
\UNTIL{$D_1 \in [cb, 2^{t+l}b-1]$}. 
\RETURN $(C, D_1, D_2)$.
\end{algorithmic}
\end{algorithm}

\begin{algorithm}
\caption{Proof of Larger Interval: $\Verify_\LI$}
\label{vlialg}
\begin{algorithmic}
\REQUIRE $\zkproof_\LI$.
\ENSURE True or false.
\RETURN $D_1 \myin [cb, 2^{t+\ell}b-1] \wedge C \meq \Hash(g^{D_1}h^{D_2}E^{-c})$.
\end{algorithmic}
\end{algorithm}

Before describing Boudot's ZKRP construction, we first need a \emph{proof with tolerance}, denoted by $\PK_\NLI = \{x, r : E = g^x h^r \wedge x \in [a-\theta, b+\theta] \}$, where $\theta = 2^{t + \ell + 1} \sqrt{b-a}$, as shown in Algorithms~\ref{pnlialg} and~\ref{vnlialg}. 

\begin{algorithm}
\caption{Proof with Tolerance: $\Prove_\NLI$}
\label{pnlialg}
\begin{algorithmic}
\REQUIRE $x, r, E$.
\ENSURE $\zkproof_\NLI$.
\STATE Compute the proof of opening of commitment $E$.
\STATE $E_a = E/g^a \pmod{n}$,
\STATE $E_b = g^b/E \pmod{n}$,
\STATE $x_a = x - a$,
\STATE $x_b = b - x$,
\STATE Alice proves that she knows $x$, which is greater that $-\theta$.
\STATE $x_{a_1} = \lfloor \sqrt{x - a} \rfloor$,
\STATE $x_{a_2} = x_a - x_{a_1}^2$,
\STATE $x_{b_1} = \lfloor \sqrt{b - x} \rfloor$,
\STATE $x_{b_2} = x_b - x_{b_1}^2$,
\REPEAT 
\STATE $r_{a_1} \in_R [-2^s n + 1, 2^s n - 1]$,
\STATE $r_{a_2} = r - r_{a_1}$,
\UNTIL {$r_{a_2} \in [-2^s n + 1, 2^s n - 1]$}.
\STATE Choose $r_{b_1}$ and $r_{b_2}$ such that $r_{b_1} + r_{b_2} = -r$.
\STATE $E_{a_1} = g^{x_{a_1}^2}h^{r_{a_1}}$,
\STATE $E_{a_2} = g^{x_{a_2}}h^{r_{a_2}}$,
\STATE $E_{b_1} = g^{x_{b_1}^2}h^{r_{b_1}}$,
\STATE $E_{b_2} = g^{x_{b_2}}h^{r_{b_2}}$,
\STATE $\zkproof_{\S_a} = \Prove_\S(x_{a_1}, r_{a_1}, E_{a_1})$,
\STATE $\zkproof_{\S_b} = \Prove_\S(x_{b_1}, r_{b_1}, E_{b_1})$,
\STATE $\zkproof_{\LI_a} = \Prove_\LI(x_{a_2}, r_{a_2}, E_{a_2})$,
\STATE $\zkproof_{\LI_b} = \Prove_\LI(x_{b_2}, r_{b_2}, E_{b_2})$,
\RETURN $\zkproof_\NLI = (E_{a_1}, E_{a_2}, E_{b_1}, E_{b_2}, \zkproof_{\S_a}, \zkproof_{\S_b}, \zkproof_{\LI_a}, \zkproof_{\LI_b})$.
\end{algorithmic}
\end{algorithm}

\begin{algorithm}
\caption{Proof with Tolerance: $\Verify_\NLI$}
\label{vnlialg}
\begin{algorithmic}
\REQUIRE $\zkproof_\NLI$.
\ENSURE True or false. 
\IF{$E_{a_2} \meq E_a/E_{a_1}  \wedge E_{b_2} \meq E_b/E_{b_1}$}
\STATE $b_\S = \Verify_\S(\zkproof_{\S_a}) \wedge \Verify_\S(\zkproof_{\S_b}),$
\STATE $b_\LI = \Verify_\LI(\zkproof_{\LI_a}), \wedge \Verify_\LI(\zkproof_{\LI_b}),$
\RETURN $b_\S \wedge b_\LI$.
\ENDIF
\RETURN False
\end{algorithmic}
\end{algorithm}

Algorithms~\ref{psdalg} and~\ref{vsdalg} describe the ZKRP scheme proposed by Boudot~\cite{boudot} in 2001. 

\begin{algorithm}
\caption{Square Decomposition Range Proof: $\Prove_\SD$}
\label{psdalg}
\begin{algorithmic}
\REQUIRE $x, r, R$.
\ENSURE $\zkproof_\SD$.
\STATE $x' = 2^T x$,
\STATE $r' = 2^T r$,
\STATE $T = 2(t + \ell + 1) + |b - a|$,
\STATE $E' =  E^{2^T}$,
\STATE $\zkproof_\NLI = \Prove_\NLI(x', r', E')$,
\RETURN $\zkproof_\SD = (E', \zkproof_\NLI)$.
\end{algorithmic}
\end{algorithm}

\begin{algorithm}
\caption{Square Decomposition Range Proof: $\Verify_\SD$}
\label{vsdalg}
\begin{algorithmic}
\REQUIRE $\zkproof_\SD$.
\ENSURE True or false.
\IF{$E' \meq  E^{2^T}$}
\RETURN $\Verify_\NLI(\zkproof_\NLI)$.
\ENDIF
\RETURN False
\end{algorithmic}
\end{algorithm}

\subsection{Signature-based construction} 
\label{ccs08}

The idea of the protocol is that the verifier initially computes digital signatures for each element in the target set $S$. The prover then blinds this digital signature by raising it to a randomly chosen exponent $v \in \ZZ_p$, such that it is computationally infeasible to determine which element was signed. The prover uses the pairing to compute the proof, and the bilinearity of the pairing allows the verifier to check that indeed one of the elements from $S$ were initially chosen. Algorithms~\ref{ssmalg},~\ref{psmalg} and~\ref{vsmalg} show the details of the this protocol.  The scheme depends upon Boneh-Boyen digital signatures, summarized in next. 

\textbf{Boneh-Boyen~\cite{bb} signatures.} Shortly, the signer private key is given by $x \in_R \ZZ_p$ and the public key is $y = g^x$. Given message $m$, we have that the digital signature is calculated as $\sigma = g^{1/(x+m)}$, and verification is achieved by computing $e(\sigma, y g^m) \meq e(g,g)$.

Boneh-Boyen signatures are based on the $q$-Strong Diffie-Hellman assumption, described in Definition~\ref{qstrongdh}. 

\begin{algorithm}
\caption{Set Membership: $\Setup_\ZKSM$}
\label{ssmalg}
\begin{algorithmic}
\REQUIRE $g, h$ and a set $S$. 
\ENSURE $y \in \G$ and $A \in G^{|S|}$.
\STATE $x \in_R \ZZ_p$,
\STATE $y = g^x$ ,
\FOR{$i \in S$}
    \STATE $A_i = g^{\frac{1}{x+i}}$.
\ENDFOR
\RETURN $y, [A_i]$.
\end{algorithmic}
\end{algorithm}

\begin{algorithm}
\caption{Set Membership: $\Prove_\ZKSM$}
\label{psmalg}
\begin{algorithmic}
\REQUIRE $g, h$, a commitment $C$, and a set $S$. 
\ENSURE $\delta,\gamma$ such that $C = g^\delta h^\gamma$ and $\delta \in S$.
\STATE $\tau \in_R \ZZ_p$, 
\STATE $V = A_\delta^\tau$,
\STATE $s,t,m \in_R \ZZ_p$,
\STATE $a = e(V, g)^{-s}.e(g,g)^t$,
\STATE $D = g^s h^m$,
\STATE $c = \Hash(V, a, D)$,
\STATE $z_\delta = s - \delta c$, 
\STATE $z_\tau = t - \tau c$,
\STATE $z_\gamma = m - \gamma c$. 
\RETURN $\zkproof_\ZKSM = (V, a, D, z_\delta, z_\tau, z_\gamma)$.
\end{algorithmic}
\end{algorithm}

\begin{algorithm}
\caption{Set Membership: $\Verify_\ZKSM$}
\label{vsmalg}
\begin{algorithmic}
\REQUIRE $g, h$, a commitment $C$, $\zkproof_\ZKSM$. 
\ENSURE True or false.
\RETURN $D \meq C^c h^{z_\gamma} g^{z_\delta} \wedge a \meq e(V,y)^c.e(V, g)^{-z_\delta}.e(g,g)^{z_\tau}$. 
\end{algorithmic}
\end{algorithm}

\textbf{Range Proof.} In order to obtain ZKRP, we can decompose the secret $\delta$ into base $u$, as follows:

$$\delta = \sum_{0 \leq j \leq \ell}{\delta_j u^j}.$$

Therefore, if each $\delta_j$ belongs to the interval $[0,u)$, then we have that $\delta \in [0, u^\ell)$. The ZKSM algorithms can be easily adapted to carry out this computation, as shown in Algorithms~\ref{srpsmalg},~\ref{prpsmalg} and~\ref{vrpsmalg}.

\begin{algorithm}
\caption{Signature-based Range Proof: $\Setup_\ZKRP$ for interval $[0, u^\ell)$}
\label{srpsmalg}
\begin{algorithmic}
\REQUIRE $g,h,u,\ell$ and a commitment $C$.
\ENSURE $\delta,\gamma$ such that $C = g^\delta h^\gamma$ and $\delta \in [0, u^\ell)$.
\STATE $x \in_R \ZZ_p$ 
\STATE $y = g^x$
\FOR{$i \in \ZZ_u$} 
    \STATE $A_i = g^\frac{1}{x+i}$.
\ENDFOR
\RETURN $y, [A_i]$.
\end{algorithmic}
\end{algorithm}

\begin{algorithm}
\caption{Signature-based Range Proof: $\Prove_\ZKRP$ for interval $[0, u^\ell)$}
\label{prpsmalg}
\begin{algorithmic}
\REQUIRE $g,h,u,\ell$ and a commitment $C$.
\ENSURE $\delta,\gamma$ such that $C = g^\delta h^\gamma$ and $\delta \in [0, u^\ell)$.
\STATE Find $[\delta_j]$ such that $\delta = \sum_j{\delta_j u^j}$,
\STATE $\tau_j \in_R \ZZ_p$, 
\STATE Set $D$ to the identity element in $\G$. 
\FOR{$j \in \ZZ_\ell$}
    \STATE $V_j = A_{\delta_j}^{\tau_j}$, 
    \STATE $s_j, t_j, m_j \in_r \ZZ_p$,
    \STATE $a_j = e(V_j, g)^{-s_j}.e(g,g)^{t_j}$,
    \STATE $D = D g^{u^j s_j}h^{m_j}$.
\ENDFOR
\STATE $c = \Hash([V_j], a, D)$. 
\FOR{$j \in \ZZ_\ell$}
    \STATE $z_{\delta_j} = s_j - \delta_j c$, 
    \STATE $z_{\tau_j} = t_j - \tau_j c$. 
\ENDFOR
\STATE $z_\gamma = m - \gamma c$.
\RETURN $\zkproof_\ZKRP = (z_\gamma,  [z_{\delta_j}], [z_{\tau_j}])$.
\end{algorithmic}
\end{algorithm}

\begin{algorithm}
\caption{Signature-based Range Proof: $\Verify_\ZKRP$ for interval $[0, u^\ell)$}
\label{vrpsmalg}
\begin{algorithmic}
\REQUIRE $g,h,u,\ell$ and $\zkproof_\ZKRP$.
\ENSURE True or false.
\STATE Set $a$ to True.
\FOR{$j \in \ZZ_\ell$}
    \STATE $a = a \wedge (a_j \meq e(V_j,y)^c.e(V_j,g)^{-z_{\delta_j}}.e(g,g)^{z_{\tau_j}})$. 
\ENDFOR
\RETURN $D \meq C^c h^{z_\gamma} \prod_j{(u^j z_{\delta_j})} \wedge a$.
\end{algorithmic}
\end{algorithm}

In order to obtain Zero Knowledge Range Proofs for arbitrary ranges $[a,b)$ we show that $\delta \in [a, a + u^\ell)$ and $\delta \in [b - u^\ell, b)$, using 2 times the ZKRP scheme described in Algorithm~\ref{prpsmalg}. Namely, we have to prove that $\delta - b + u^\ell \in [0, u^\ell)$ and $\delta - a \in [0, u^\ell)$.

\subsection{Bulletproofs construction}
\label{bpsec}

In this section we show a detailed description of the algorithms necessary to implement the Bulletproofs ZKRP protocol. 

\textbf{Notation.} Given an array $\a \in \G^n$, we use Python notation to represent array slices:

$$\a_{[:\ell]} = [a_1, \dots, a_\ell] \in \G^\ell,$$

$$\a_{[\ell:]} = [a_{\ell+1}, \dots, a_n] \in \G^{n - \ell}$$

Given $k \in \G$, we denote the vector containing the powers of $k$ by 

$$\k^n = [1, k, k^2, \dots, k^{n-1}].$$

Given $\g = [g_1, \dots, g_n] \in \G^n$ and $\a \in \Z_p^n$, we define $\g^\a$ as follows:

$$\g^\a = \prod_{i=1}^n g_i^{a_i}.$$

Given $c \in \Z_p$, notation $\b = c.\a \in \Z_p^n$ is a vector such that $b_i = c.a_i$. Also, $\a \circ \b = (a_1 b_1, \dots, a_n b_n)$ is the Hadamard product. The vector polynomial $p(X) = \sum_{i=0}^n \p_i X^i \in \Z_p^n[X]$, where each coefficient $\p_i$ is a vector in $\Z_p^n$. The inner product of such polynomials is given by 

\begin{equation}
\label{ipdef}
\langle \l(X), \rr(X) \rangle = \sum_{i=0}^d \sum_{j=0}^i \langle \l_i, \rr_j \rangle X^{i+j} \in \Z_p[X].
\end{equation}

\subsubsection{Setup}

Many ZKRP constructions depend on a \textbf{trusted setup}. Shortly, the parameters necessary to generate and verify the underlying zero knowledge proofs must be computed by a trusted party, because if such parameters are generated using a \emph{trapdoor}, then this trapdoor could be used to subvert the protocol, allowing to generate money out of thin air.  

In order to avoid the trusted setup, Bulletproofs use the \emph{Nothing Up My Sleeve} (NUMS) strategy, where a hash function~\cite{maptogroup} is utilized to compute the generators that will be necessary for the Pedersen commitments, as described in Algorithm~\ref{setupalg}, which describes the specific case where the subjacent elliptic curve is given by Koblitz curve \texttt{secp256k1}~\cite{sec2,koblitz}. 

\begin{algorithm}
\begin{algorithmic}
\REQUIRE The input string $m$, and the field prime modulus $p$. 
\ENSURE An elliptic curve point if successful or some error. 
	\STATE $i = 0$.
	\WHILE{$i < 256$}
		\STATE $x = \Hash(m, i)$.
		\STATE $\rhs = x^3 + 7 \pmod{p}$.
		\IF{$\rhs$ is a square $\pmod{p}$}
			\STATE $y = \sqrt{\rhs} \pmod{p}$.
			\IF{$(x, y)$ is not the point at infinity}
				\RETURN $(x,y)$.
			\ENDIF
		\ENDIF
		\STATE $i = i + 1$.
	\ENDWHILE
	\RETURN ``Can not map to group''.
\end{algorithmic}
	\caption{Nothing Up My Sleeve: $\MapToGroup$}
\label{setupalg}
\end{algorithm}

\begin{algorithm}
\begin{algorithmic}
\REQUIRE The elliptic curve public generator $g \in \G$ and an integer $n$. 
\ENSURE The set of generators $(g, h, \g, \h)$.
	\STATE Compute $h = \MapToGroup(\mathrm{'some \ public \ string'}, p)$.
	\STATE $i = 0$.
	\WHILE{$i < n$}
		\STATE $c \in_R \Z_p,$
		\STATE $d \in_R \Z_p,$
		\STATE $\g[i] = c.G,$
		\STATE $\h[i] = d.G,$
		\STATE $i = i + 1$.
	\ENDWHILE
	\RETURN $(g, h, \g, \h)$.
\end{algorithmic}
	\caption{Compute Generators: $\CompGen$}
\label{compgenalg}
\end{algorithm}

\begin{algorithm}
\begin{algorithmic}
\REQUIRE The set of generators $(g, h, \g, \h)$. 
\ENSURE $\params_\IP$. 
	\STATE $u = \MapToGroup(\mathrm{'some \ other \ public \ string'}, p)$. 
	\RETURN $\params_\IP = (g, h, \g, \h, u)$.
\end{algorithmic}
	\caption{$\Setup_\IP$}
\label{setupipalg}
\end{algorithm}

\begin{algorithm}
\begin{algorithmic}
\REQUIRE The input interval $[a, b)$ and the field modulus $p$. 
\ENSURE $\params_\RP$. 
	\IF{$b$ is not a power of 2}
		\RETURN ``$b$ must be a power of 2''.
	\ELSE
		\STATE $n = \log_2{b},$
		\STATE $(g, h, \g, \h) = \CompGen(g, n),$
		\STATE $\params_\IP = \Setup_\IP(g, h, \g, \h),$ 
		\STATE $\params_\RP = (\params_\IP, n),$
		\RETURN $\params_\RP$.
	\ENDIF
\end{algorithmic}
	\caption{$\Setup_\RP$}
\label{setuprpalg}
\end{algorithm}

\subsubsection{Inner product argument} 
\label{ipsec}

In this section we present the main building block of Bulletproofs, which is the inner product argument. In summary, using this ZKP protocol the prover convinces a verifier that she knows vectors whose inner product is equal to a determined public value. First we describe the initialization procedure in Algorithm~\ref{bp1alg}. Afterwards we present the main protocol, given by Algorithm~\ref{bp2alg}.

\begin{algorithm}
\begin{algorithmic}
\REQUIRE $(\params_\IP, \a, \b)$. 
\ENSURE The commitment $P$.
	\STATE Compute $P = \g^\a \h^\b \in \G$.
	\RETURN $P$.
\end{algorithmic}
	\caption{Vector Commitment: $\Commit_\IP$}
\label{calg}
\end{algorithm}

\begin{algorithm}
\begin{algorithmic}
\REQUIRE $(\params_\IP, \commit_\IP, c, \a, \b)$. 
\ENSURE $\zkproof_\IP$.
	\STATE $x = \Hash(\g, \h, P, c) \in \Z_p^\star$
	\STATE Compute $P' = u^{x.c} P$.
	\STATE Allocate arrays $\l, \rr \in \G^n$.
	\STATE $\Recursion(\g, \h, P', u^x, \a, \b, \l, \rr)$.
	\STATE $\zkproof_\IP = (\g, \h, P', u^x, \a, \b, \l, \rr)$.
	\RETURN $\zkproof_\IP$.
\end{algorithmic}
	\caption{Proof of Inner Product: $\Prove_\IP$}
\label{bp1alg}
\end{algorithm}

\begin{algorithm}
\caption{Proof of Inner Product: $\Recursion$}
\label{bp2alg}
\begin{algorithmic}
\REQUIRE $(\g, \h, P, u, a, b, \l, \rr)$.
\ENSURE $(\g, \h, P, u, a, b, \l, \rr)$.
	\STATE $x = \Hash(\g, \h, P, c) \in \Z_p^\star$.
	\STATE Compute $P' = u^{x.c} P$.
	\IF {$n = 1$}
		\RETURN $(\g, \h, P, u, a, b, \l, \rr)$.
	\ELSE 
		\STATE $n' = \frac{n}{2},$
		\STATE $c_L = \langle \a_{[:n']}, \b_{[n':]} \rangle \in \Z_p,$
		\STATE $c_R = \langle \a_{[n':]}, \b_{[:n']} \rangle \in \Z_p,$
		\STATE $L = \g_{[n':]}^{\a_{[:n']}} \h_{[:n']}^{\b_{[n':]}} u^{c_L} \in \G,$ 
		\STATE $R = \g_{[:n']}^{\a_{[n':]}} \h_{[n':]}^{\b_{[:n']}} u^{c_R} \in \G$.
		\STATE Append $L, R$ to $\l, \rr$, respectively.
		\STATE $x = \Hash(L, R),$
		\STATE $\g' = \g_{[:n']}^{x^{-1}} \g_{[n':]}^{x} \in \G^{n'},$
		\STATE $\h' = \g_{[:n']}^{x} \g_{[n':]}^{x^{-1}} \in \G^{n'},$
		\STATE $P' = L^{x^2} P R^{x^{-2}} \in \G,$
		\STATE $\a' = \a_{[:n']} x +\a_{[n':]} x^{-1} \in \Z_p^{n'},$
		\STATE $\b' = \b_{[:n']} x^{-1} +\b_{[n':]} x \in \Z_p^{n'}.$
		\STATE Recursively run $\Recursion$ on input $(\g',\h', P', u, \a', \b', \l, \rr)$. 
	\ENDIF
\end{algorithmic}
\end{algorithm}

\begin{algorithm}
\caption{Proof of Inner Product: $\Verify_\IP$}
\label{vbp2alg}
\begin{algorithmic}
\REQUIRE $\params_\IP, \commit_\IP, \zkproof_\IP$. 
\ENSURE True or false.
\STATE $i = 0$.
	\WHILE{$i < \log{n}$}
	\STATE $n' = \frac{n}{2},$
	\STATE $x = \Hash(\l[i], \rr[i]),$
	\STATE $\g' = \g_{[:n']}^{x^{-1}} \g_{[n':]}^{x} \in \G^{n'},$
	\STATE $\h' = \g_{[:n']}^{x} \g_{[n':]}^{x^{-1}} \in \G^{n'},$
	\STATE $P' = L^{x^2} P R^{x^{-2}} \in \G,$
	\STATE $i = i + 1$.
\ENDWHILE
\STATE The verifier computes $c = a.b$ and accepts if $P = g^a h^b u^c$.
\end{algorithmic}
\end{algorithm}

The fact that Bulletproofs allows to halve the size of the problem in each level of the recursion in Algorithm~\ref{bp2alg} means that it is possible to obtain logarithmic proof size. 

\subsubsection{Range proof argument}

Given a secret value $v$, if we want to prove it belongs to the interval $[0, 2^n)$, then we do the following:

\begin{itemize}
	\item Prove that $\a_L \in \{0,1\}^n$ is the bit-decomposition of $v$. In other words, we show that 

		$$\langle \a_L, \2^n \rangle = v.$$ 

	\item Define $\a_R$ as the component-wise complement of $\a_L$, what means that, for every $i \in [0,n]$, if the $i$-th bit of $\a_L$ is $0$, then the $i$-th bit of $\a_R$ is equal to $1$. Conversely, if the $i$-th bit of $\a_L$ is $1$, then the $i$-th bit of $\a_R$ is equal to $0$. Equivalently, this condition can be shortly described by Equations~\ref{compl1} and~\ref{compl2}.  

\begin{equation}
\label{compl1}
\a_L \circ \a_R = \0^n,
\end{equation}

\begin{equation}
\label{compl2}
\a_R = \a_L - 1^n \pmod{2}.
\end{equation}

In order to prove that $\a_L$ and $\a_R$ satisfy both relations, we can randomly choose $y \in \Z_p$ and compute:

$$\langle \a_L, \a_R \circ \y^n \rangle = 0,$$

$$\langle \a_L - \1^n - \a_R, \y^n \rangle = 0.$$

These two equations can be combined into a single inner product, by randomly choosing $z \in \Z_p$, and computing

\begin{equation}
\label{eq1}
\langle \a_L - z.\1^n, \y^n \circ (\a_R + z.\1^n) + z^2.\2^n \rangle = z^2 v + \delta(y,z),
\end{equation}

where $\delta(y,z) = (z - z^2) \langle \1^n, \y^n \rangle - z^3 \langle \1^n, \2^n  \rangle \in \Z_p$. 
\end{itemize}

If the prover could send the vectors in Equation~\ref{eq1}, then the verifier would be able to check the inner product himself. However, this vector reveals information about $\a_L$, therefore revealing bits of the secret value $v$. To solve this problem the prover randomly chooses vectors $\s_L$ and $\s_R$ in order to blind $\a_L$ and $\a_R$, respectively. Consider the following polynomials:

$$l[X] = \a_L - z.1^n + s_L.X \in \Z_p^n,$$

$$r[X] = \y^n \circ (\a_R + z.1^n + \s_R.X) + z^2 2^n \in \Z_p^n,$$

$$t[X] = \langle l[X], r[X] \rangle = t_0 + t_1.X + t_2.X^2,$$
where the above inner product is computed as defined in Equation~\ref{ipdef}.

Note that the constant terms of $l[X]$ and $r[X]$ correspond to the vectors in Equation~\ref{eq1}. Therefore if the prover publishes $l[x]$ and $r[x]$ for a specific $x \in \Z_p$, then we have that terms $\s_L$ and $\s_R$ ensure no information about $\a_L$ and $\a_R$ is revealed.  

Explicitly, we have that 

\begin{equation}
\label{eqt1}
t_1 = \langle \a_L - z.\1^n, \y^n.\s_R \rangle + \langle \s_L, \y^n.(\a_R + z.\1^n) \rangle,
\end{equation}
and

\begin{equation}
\label{eqt2}
t_2 = \langle \s_L, \y^n.\s_R \rangle.
\end{equation}

\begin{algorithm}
\begin{algorithmic}
\REQUIRE $\params_\RP, v$.
\ENSURE $\zkproof_\RP$.
\STATE $\gamma \in_R \Z_p,$
\STATE $V = g^v h^\gamma \in \G,$
\STATE $\a_L \in \{0,1\}^n$ such that $\langle \a_L, \2^n \rangle = v,$ 
\STATE $\a_R = \a_L - \1^n \in \Z_p^n,$
\STATE $\alpha \in_R \Z_p,$
\STATE $A = h^\alpha \g^{\a_L} \h^{\a_R} \in \G,$
\STATE $s_L, s_R \in_R \Z_p^n,$
\STATE $\rho \in_R \Z_p,$
\STATE $S = h^\rho \g^{s_L} \h^{s_R} \in \G,$
\STATE $y = \Hash(A, S) \in \Z_p^\star,$
\STATE $z = \Hash(A, S, y) \in \Z_p^\star,$
\STATE $\tau_1, \tau_2 \in_R \Z_p,$
\STATE $T_1 = g^{t_1} h^{\tau_1} \in \G,$
\STATE $T_2 = g^{t_2} h^{\tau_2} \in \G,$
\STATE $x =\Hash(T_1, T_2) \in \Z_p^\star,$
\STATE $\l = l(X) = \a_L - z 1^n + s_L X \in \Z_p^n,$
\STATE $\rr = r(X) = \y^n \circ (\a_R + z 1^n + \s_R X) + z^2 2^n \in \Z_p^n,$
\STATE $\hat{t} = \langle \l, \rr \rangle \in \Z_p,$
\STATE $\tau_x = \tau_2 x^2 + \tau_1 x + z^2 \gamma \in \Z_p,$
\STATE $\mu = \alpha + \rho x \in \Z_p,$
\STATE $\commit_\IP = \Commit_\IP(\params_\IP, \l, \rr),$
\STATE $\zkproof_\IP = \Prove_\IP(\params_\IP, \commit_\IP, \hat{t}, \l, \rr),$  
\STATE $\zkproof_\RP = (\tau_x, \mu, \hat{t}, V, A, S, T_1, T_2, \commit_\IP, \zkproof_\IP)$.	
\RETURN $\zkproof_\RP$.	
\end{algorithmic}
	\caption{Bulletproofs: $\Prove_\RP$}
\label{prpalg}
\end{algorithm}

\begin{algorithm}
\begin{algorithmic}
\REQUIRE $\params_\RP, \zkproof_\RP$.
\ENSURE True or false.
\STATE $y = \Hash(A, S) \in \Z_p^\star,$
\STATE $z = \Hash(A, S, y) \in \Z_p^\star,$
\STATE $x =\Hash(T_1, T_2) \in \Z_p^\star,$
\STATE $h_i =  h_i^{y^{-i+1}} \in \G, \forall i \in [1,n],$
\STATE $P_l = P.h^{\mu},$
\STATE $P_r = A.S^x.\g^{-z}.(\h')^{z.\y^n + z^2.\2^n} \in \G,$
\STATE $\out_1 = (P_l \meq P_r),$
\STATE $\out_2 = (g^{\hat{t}} h^{\tau_x} \meq V^{z^2}.g^{\delta(y,z)}.T_1^x.T_2^{x^2}),$
\STATE $\out_3 = \Verify_\IP(\zkproof_\IP),$
\RETURN $\out_1 \wedge \out_2 \wedge \out_3$. 
\end{algorithmic}
	\caption{Bulletproofs: $\Verify_\RP$}
\label{vrpalg}
\end{algorithm}

In order to make Bulletproofs non-interactive using the Fiat-Shamir heuristic. Concretely, we compute $x = \Hash(T_1, T_2)$, $y = \Hash(A, S)$, and $z = \Hash(A, S, y)$ in Algorithms~\ref{prpalg} and~\ref{vrpalg}.

\subsubsection{Optimizations}

The algorithms described in last section can be optimized in two ways, as follows:

\begin{itemize}
\item \textbf{Multi-exponentiation.} In the inner-product argument presented in Section~\ref{ipsec} it is required to computed many exponentiations, which is an expensive operation. For instance, in the $k$-th round of the protocol we must perform $\frac{n}{2^{k-1}}$ exponentiations, thus in total we must execute $4n$ exponentiations. It is possible to reduce this number to a single \emph{multi-exponentiation} of size $2n$ by postponing these computations to the last round.
	
Concretely, given $\g = [g_1, \dots, g_n]$, we have that it is possible to compute $g$ an $h$, the generators obtained in last round, by using the following expressions:

$$g = \prod_{i=1}^n g_i^{s_i} \in \G, $$

$$h = \prod_{i=1}^n h_i^{1/s_i} \in \G,$$ 
where 

$$s_i = \prod_{j=1}^{\log_2{n}} x_j^{b(i,j)}$$
and 
$$b(i,j) = \begin{cases}
	1, & \text{if the $j$-th bit of $i-1$ is 1}\\
	-1, & \text{otherwise}
\end{cases}$$

Therefore, verification can be performed by 

$$\g^{a.\s}.\h^{b.s^{-1}}.u^{a.b} \meq P.\prod_{j=1}^{\log_2{n}} L_j^{x_j^2}.R_j^{x_j^{-2}}.$$

\item \textbf{Aggregation.} If multiple range proofs use the same underlying interval, then it is possible to aggregate them into one single ZKRP. Using this optimization, we have that new proofs can be added by only increasing the total size of the proof by a logarithmic factor. Consider we want to aggregate $m$ range proofs. Then, while the naive strategy would lead us to a proof whose size is $m$ times larger, this aggregation procedure in Bulletproofs allows the proof to grow only by a factor of $2\log_2{m}$.

In practice, applications like Confidential Transactions~\cite{ct}, Mimblewimble~\cite{mimblewimble} and Provisions~\cite{provisions} would benefit a lot from the utilization of aggregation, because indeed such applications must execute many ZKRPs over the same interval. 
\end{itemize}

\section{Implementation}
\label{impl}

We implemented the constructions described in Sections~\ref{boudot},~\ref{ccs08} and~\ref{bpsec}. The scheme based on square decomposition, i.e. Boudot's construction, was implemented in Java and Solidity, while the signature-based scheme and Bulletproofs were implemented in Golang and they were based on libsecp256k1 library, available in Go-Ethereum. We used BN128 pairing-friendly elliptic curves, thus accomplishing 128 bits of security. The performance is summarized in Table~\ref{restab}, and the measurement was carried out in a computer with a 64-bit Intel i5-6300U 2.40GHz CPU, 16 GB of RAM and Ubuntu 18.04. The implementation is available on Github~\cite{impl} and is a proof of concept, thus it should not be used in production without first spending the effort to review it where necessary.   

Optimal values for $u$ and $\ell$ can be calculated as described in the original paper~\cite{ccs08}. We used $u = 57$ and $\ell = 5$ for the interval $[347184000, 599644800)$, obtaining communication complexity equal to 30976 bits, while the previous work, based on Boudot's proposal~\cite{boudot}, has 48946 bits. 

\begin{table}[h]
	\centering
	\begin{threeparttable}
	\begin{tabular}{| c | c | c | c |}
		\hline
		Scheme & Setup (ms) & Prove (ms) & Verify (ms) \\ \hline
		\cite{boudot} & 331.41 & 579.32 & 851.89 \\ \hline
		\cite{ccs08} & 31.78 & 70.18 & 98.95 \\ \hline
		\cite{bulletproofs} & 13.11 & 96.25 & 51.86 \\ \hline
		\cite{bulletproofs}~\tnote{a} & 17.20  & 22.38 & 3.27 \\ \hline
	\end{tabular}
	\begin{tablenotes}
	\item[a] optimized implementation 
	\end{tablenotes}
	\end{threeparttable}
\caption{Time complexity}
\label{restab}
\end{table}

A more detailed comparison is presented in next section, showing the performance data for different range sizes. 

\section{Comparison}
\label{comp}

In this section we compare the schemes presented in Section~\ref{main} with respect to proof size and the complexity of $\Prove$ and $\Verify$ algorithms. We chose the most efficient proposal from each different strategy in order to do the comparison. Namely, we used the proposal by Lipmaa et al~\cite{mbase03} to represent the multi-base solution; the proposal by Lipmaa~\cite{lip03} to represent the square decomposition strategy; the scheme by Camenisch et al~\cite{ccs08} for the signature-based implementation; and the Bulletproofs construction. This comparison extends the work by Canard et al~\cite{survey1} to include Bulletproofs's performance.

Compared to other proposals in the literature, we found that for very big intervals, the best strategy is to use the square decomposition, as for example occurs in the construction by Boudot~\cite{boudot}, since verification doesn't depend on the size of the secret. However, it is important to remark that finding the decomposition into squares consumes a reasonable amount of computational resources, what makes the Prover's algorithm somewhat inefficient. On the other hand, for small secrets, Schoenmakers's strategy~\cite{schoenmakers} is the most efficient scheme with respect to the $\Prove$ algorithm. 

Although we implemented the scheme described in Section~\ref{boudot}, namely Boudot's proposal~\cite{boudot}, we have that its performance is very similar when compared to Lipmaa's~\cite{lip03} and Groth's~\cite{gro05} constructions. 

\begin{figure}
  \caption{Proofs size}
  \label{cfig}
  \centering
    \includegraphics[width=0.45\textwidth]{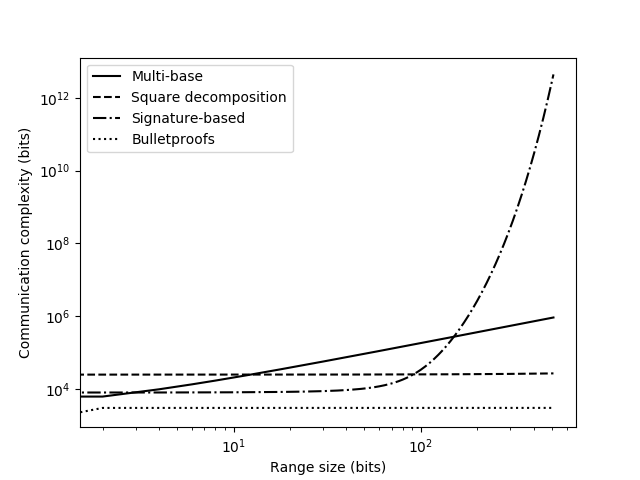}
\end{figure}

\begin{figure}
  \caption{Prover complexity}
  \label{pfig}
  \centering
    \includegraphics[width=0.45\textwidth]{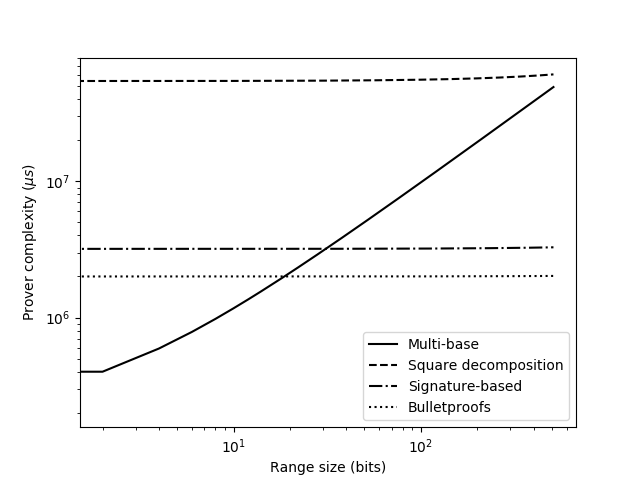}
\end{figure}

\begin{figure}
  \caption{Verifier complexity}
  \label{vfig}
  \centering
    \includegraphics[width=0.45\textwidth]{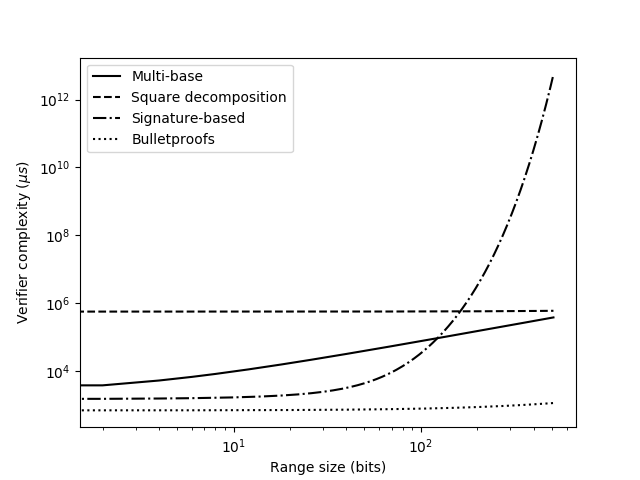}
\end{figure}

In Figures~\ref{cfig},~\ref{pfig} and~\ref{vfig} we represent in the horizontal axis the bit-length of $b$, where $b$ is the largest element from the subjacent range $[a,b]$ used for the zero knowledge range proof scheme. 

It is possible to conclude that in general Bulletproofs offers the best performance, but depending on the requirements of the underlying chosen use case, it may be possible that other strategies offer better advantages. For DLT applications we have that the proof size and the verifier's complexity are more important metrics than the prover's complexity, what means that indeed Bulletproofs seems to be the best approach to implement a ZKRP protocol. 

\section{Related work and final remarks}
\label{fmsec}

In this document we described in detail the construction of ZKRP and ZKSM protocols, which were implemented over Go-Ethereum library. Another way to obtain Zero Knowledge Set Membership protocols is by using \emph{cryptographic accumulators}~\cite{accumulators1,accumulators2,accumulators3}. Also, the underlying digital signature scheme used, namely Boneh-Boyen signatures, can be replaced and the construction presented here can be adapted to use the digital signature proposed by Camenisch and Lysyanskaya~\cite{CL}. Nevertheless, both modifications would make it necessary to assume hardness of the strong RSA assumption.  

In the context of DLT applications, it is possible to use Zero Knowledge Set Membership to validate user information without revealing it. A possible scenario is to perform KYC operations. For example, it would be possible to validate that the country of residence of a user is one belonging to the European Union, without revealing which country. In the case of Zero Knowledge Range Proofs, a commonly mentioned application is validating that someone is over 18 and thus is allowed to use a certain service, without revealing the age. In Section~\ref{int} we discussed several other applications, like reputation systems and AML or CRS compliance. 

Recent breakthroughs in cryptography permit us to construct new protocols and achieve \emph{privacy on demand}. These new cryptographic algorithms can be ultimately considered as tools that can be reused in different problems. Therefore ING is following the steps to build the knowledge that is necessary in order to construct a \emph{toolbox} to deal with the above-mentioned complex problems.  

As a future work, we will integrate this implementation to Ethereum, such that ZKSM can be used in a smart contract. In order to do that, it would be interesting to rewrite the $\Verify$ algorithm in Solidity, avoiding the necessity of using our modified Go-Ethereum client. Also, we will research other ZKP protocols that may be used to enhance privacy on DLT and blockchain. 

Finally, an important research topic is the construction of post-quantum zero knowledge proofs. Recently, Beno\^it Libert et al~\cite{pqrp} proposed a construction of ZKRP based on lattices. However, the proof size is 3.54 MB for secret whose size is $2^{1000}$. Although the secret is huge, the size of the proof can't be made considerably smaller when the secrets is smaller. Hence, optimizing this construction would allow to reduce the gap existing between conventional schemes and quantum-resistant ones. 

\bibliography{refs}

\end{document}